\documentclass[11pt, onecolumn, a4paper]{article}

\usepackage{times}
\usepackage{setspace}
\usepackage{textcomp}
\usepackage{amsmath}
\usepackage{graphicx}
\usepackage{amssymb}
\usepackage{cite}
\usepackage{ulem}
\usepackage{color}

\graphicspath{{./figs/}}

\doublespace

\begin{document}

\title{Four-dimensional light shaping: manipulating ultrafast spatio-temporal foci in space and time}

\author{Bangshan Sun$^{1,3}$, Patrick S. Salter$^3$, Clemens Roider$^{1,4}$, Alexander Jesacher$^{2,4}$, Johannes Strauss$^{1,2}$,\\
 Johannes Heberle$^{1,2}$, Michael Schmidt$^{1,2}$ and Martin J. Booth$^{2,3,5,*}$}



\maketitle

\begin{center}
\small
\noindent $^1$Institute of Photonic Technologies, Friedrich-Alexander-University Erlangen-Nuremberg,\\ Konrad-Zuse-Strasse 3/5, 91052 Erlangen, Germany\\
$^2$Graduate School in Advanced Optical Technologies (SAOT), Friedrich-Alexander University Erlangen-Nuremberg,\\ Paul-Gordan-Strasse 6, 91052 Erlangen, Germany\\
$^3$Department of Engineering Science, University of Oxford, Parks Road, Oxford,OX1 3PJ, UK\\
$^4$Division of Biomedical Physics, Innsbruck Medical University, Mullerstrasse 44, 6020 Innsbruck, Austria\\
$^5$Centre for Neural Circuits and Behaviour, University of Oxford, Mansfield Road, Oxford, OX1 3SR, UK

\noindent$^*$Email: martin.booth@eng.ox.ac.uk \\
\end{center}

\begin{abstract} Spectral dispersion of ultrashort pulses allows simultaneous focusing of light in both space and time creating so-called spatio-temporal foci. Such space-time coupling may be combined with existing holographic techniques to give a further dimension of control when generating focal light fields. It is shown that a phase-only hologram placed in the pupil plane of an objective and illuminated by a spatially chirped ultrashort pulse can be used to generate three dimensional arrays of spatio-temporally focused spots. Exploiting the pulse front tilt generated at focus when applying simultaneous spatial and temporal focusing (SSTF), it is possible to overlap neighbouring foci in time to create a smooth intensity distribution. The resulting light field displays a high level of axial confinement, with experimental demonstrations given through two-photon microscopy and non-linear laser fabrication of glass.

\end{abstract}


\normalsize
\section*{INTRODUCTION}

The control of light in the focal region of a microscope objective lens is important in many different applications ranging from imaging~\cite{ritsch2011}, to optical trapping~\cite{grier2002} and laser fabrication~\cite{Sugioka2014}. Computer generated holography has become a widespread tool to provide such control, particularly as it has developed in parallel with the increasing performance of liquid crystal spatial light modulators, to the extent that dynamic updating of complex light fields is now achievable in real time~\cite{preece2009}. However, the development of control of such complex focal fields has mostly been confined only to consideration of the three spatial dimensions.

When focusing an ultrashort pulse, a further degree of freedom may be achieved. Dispersion of the spectral components of the light can stretch the pulse temporally, after which it may be compressed again to reconstitute the original short pulse. This process limits the interaction range of a non-linear process, such as two-photon fluorescence~\cite{xu05, Oron2005, Durfee2014}. Such temporal focusing may be combined with spatial focusing, for example using a grating followed by a positive lens, but in general the two phenomena can be decoupled. Previously such simultaneous spatio-temporal focusing (SSTF) had been used to generate wide-field focal intensity distributions with high uniformity and strong axial confinement, enabling rapid multiphoton imaging~\cite{Choi2013, Lien2014, Dana2014, Song2015}, with potential for use even deep inside scattering tissue~\cite{Shoham2013}. The spectral spread prior to focus also alleviates problems related to non-linear propagation, such as filamentation, which can prove disastrous for laser processing through thick dielectrics~\cite{Vitek2010a,Kammel2014}.  In addition, SSTF offers several further levels of control over the ultrafast laser focus, including a controllable aspect ratio~\cite{He2010,He2011} and an adjustable pulse front tilt (PFT)~\cite{Vitek2010,Block2014,Zhang2014} leading to applications such as processing glass for microfluidics~\cite{Vitek2010a, He2010} and biomedical tissue ablation~\cite{Stoian2013}.

Parallelization is already a common concept using conventional focusing for increasing process speed in applications such as multiphoton microscopy~\cite{Poland2014} and laser fabrication~\cite{Waller2013, Lin2013, Hasagawa2016}. SSTF has further advantages when generating such complex focal distributions. By shaping the light through holography~\cite{Papagiakoumou2013, Hernandez2016}, projection~\cite{Papagiakoumou2008, Kim2010,Li2012} or generalised phase contrast~\cite{Papagiakoumou2010} prior to the dispersive element, a patterned SSTF focus may be generated. Such continuous intensity distributions typically display poor axial resolution, but including SSTF in the optical system restores axial confinement at the focus. In this paper, we demonstrate that it is also possible to holographically shape  the light after spectral dispersion of the pulse. We are able to create two and three dimensional arrays of SSTF spots using a phase-only hologram placed in the spatially chirped beam. In fact, using a hologram in this configuration gives a further level of control to the light field. The inherent pulse front tilt associated with SSTF spots can be exploited to  overlap neighbouring foci, such that multiple spots from a single pulse may overlap in space but be separate in time. Thus, the conventional limitations on holographic spot spacing due to mutual interference can be overcome through simultaneous control of the intensity distribution in four dimensions: three of space and one of time.

\section*{MATERIALS AND METHODS}

\subsection*{Holography in a spatially chirped beam}

How can holographic elements be used to adjust the focus in the time dimension?  It is already known that applying group velocity dispersion (GVD) to an ultrafast beam axially scans a SSTF focus~\cite{Durst2006}. This can be simply understood if the GVD corresponds to a quadratic spectral phase (that is the phase shift $\Psi$ at a frequency $\omega$ is given by $\Psi(\omega) \propto \omega^2 $).  In the pupil plane of the objective in a SSTF system, there is a uniform spatial chirp (such that $r(\omega)\propto \omega$, where $r$ is the radial position in the pupil of a spectral component). Hence, $\Psi(r)\propto r^2$, which shows that the GVD is equivalent to applying a quadratic phase in the objective pupil, which in the paraxial regime shifts the focus along the optical axis~\cite{Lesham2014, Hernandez2016, Zurauskas2017}. Recent work has also shown that applying various phase-based aberrations in the pupil plane of an objective leads to some familiar focal distortion with reference to conventional focusing systems~\cite{Sun2014, Chang2014, Greco2016}. So what is the link between these spatial manipulations and temporal control?

It is instructive here to look at the application of a linear phase gradient to a spatially chirped beam in the pupil of an objective lens (in other words, $\Psi(r)\propto r$). Figure~\ref{instruction}(a) shows the space-time ($x$, $t$) intensity variation of an example SSTF focus. There is a strong pulse front tilt (PFT), which is characteristic of SSTF~\cite{Durfee2012}. As should be expected, the linear phase gradient causes a transverse shift of the focus. However, this phase gradient is also equivalent to a spectral phase $\Psi(\omega) \propto \omega$. So, from the Fourier shift theorem, one should also expect a temporal delay in the focus.  This dual manifestation is apparent in the plot of Figure~\ref{instruction}(b), where we see that the combination of the lateral shift and the PFT leads to a temporal delay on axis. Thus we can see that it is indeed possible to translate a single SSTF focus in both space and time with an appropriate phase applied to the pupil plane of the lens.

Since it is possible to translate a SSTF focus in three dimensions using a phase pattern placed in the spatially chirped beam, it is an obvious extension that a combination of gratings to direct foci to different locations within the focal region should enable generation of multiple SSTF spots. The principle is shown in the sketch included in Figure~\ref{instruction}(c). The pattern displayed on the SLM modulates the phases of all spectral components, and each spectral component is diffracted by the unique local region of the hologram and propagates along several different beam paths into the focal region. The net effect of this is that there are several parts of the focal region where all spectral components overlap with the correct phase to form spatiotemporal foci. As a result, each spatiotemporal focal spot is expected to contain contributions from all spectral components across the whole pupil, confirming the simultaneous spatial and temporal focusing property for all the spots.  Furthermore, it should be possible within certain practical bounds to introduce desired spatial and temporal shifts to the different foci.

\subsection*{Experimental system}

\begin{figure}[htb]
\centering\includegraphics[width=15cm]{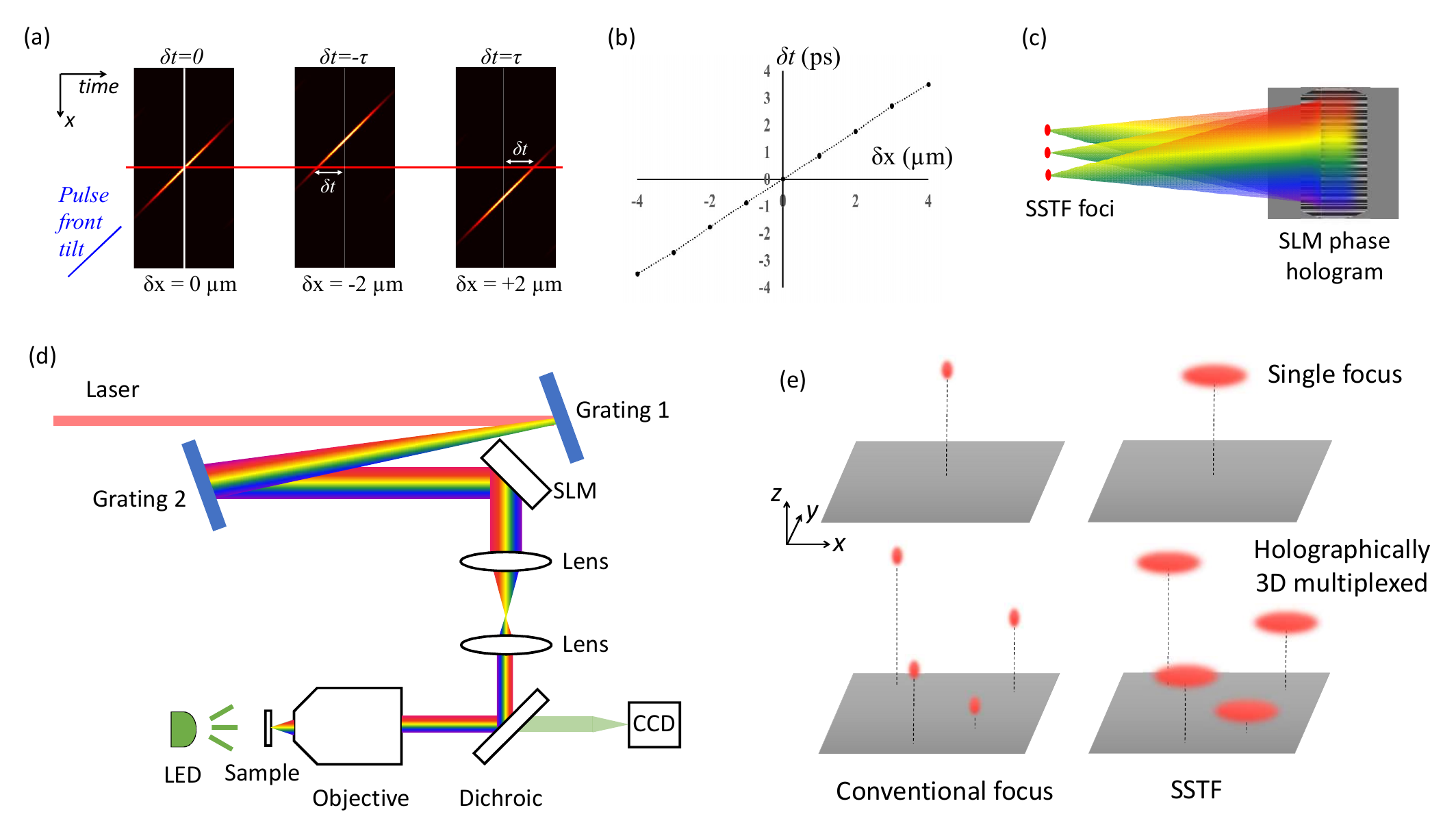}
\caption{(a) Effect of a linear phase gradient on the SSTF focus. Images show simulated results of space-time (x, t) intensity variations of temporal foci. Left to right: no phase gradient, linear phase causing negative shift, linear phase causing positive shift. (b) Plot of lateral spatial shift versus on-axis temporal delay for a SSTF focus, with a varying magnitude of linear spectral phase applied. (c) Sketch illustrating the principle for creating a three-dimensional multiple SSTF array. (d) Experimental system for four-dimensional light shaping (full diagram in Supplementary information Figure~S1).  (e) Sketch representing the relationship between ``conventional focus" and ``SSTF", as well as ``single focus" and ``holographic 3D multiplexed".}
\label{instruction}
\end{figure}

The experimental system is shown in Figure~\ref{instruction}(d). The ultrafast laser was a OneFive Origami XP, with wavelength at 1030~nm, pulse duration of 350~fs and tunable repetition rate from 50~kHz to 1~MHz. The beam was spectrally spread using two blazed gratings (Thorlabs GR25-0610, 600 grooves/mm, blazed for 1~$\mathrm{\mu}$m wavelength light). The incident angle for both gratings was aligned to approach the Littrow Configuration (17$^{\circ}$27$'$), giving a first order diffraction efficiency of 85\% per grating. The spectrally dispersed beam was then directed onto a phase only liquid crystal  SLM (Hamamatsu X11840). The full width half maximum (FWHM) of the  beam in the unchirped direction $D_x $ at the SLM was only around 3.1~mm, limiting the active area. This dimension $D_x$ determines the monochromatic beam size at the pupil, which can be adjusted to control the lateral ellipticity of each SSTF focus~\cite{He2010,He2011}. A smaller $D_x $ extends the dimensions of the spatio-temporal focus axially and laterally (in a direction orthogonal to the spatial chirp)~\cite{Sun2014}. A 4f system was used to image the  SLM into the pupil of the objective lens (Zeiss EC Epiplan 50$\times$, 0.75~NA and a pupil diameter of 4.95~mm). A 3D translation stage (Thorlabs 6-Axis NanoMax, MAX607/M) was used to control the motion of the sample. A LED-illuminated  widefield optical microscope was integrated to image the focal region.

\subsection*{Hologram design}

\begin{figure}[htb]
\centering\includegraphics[width=12cm]{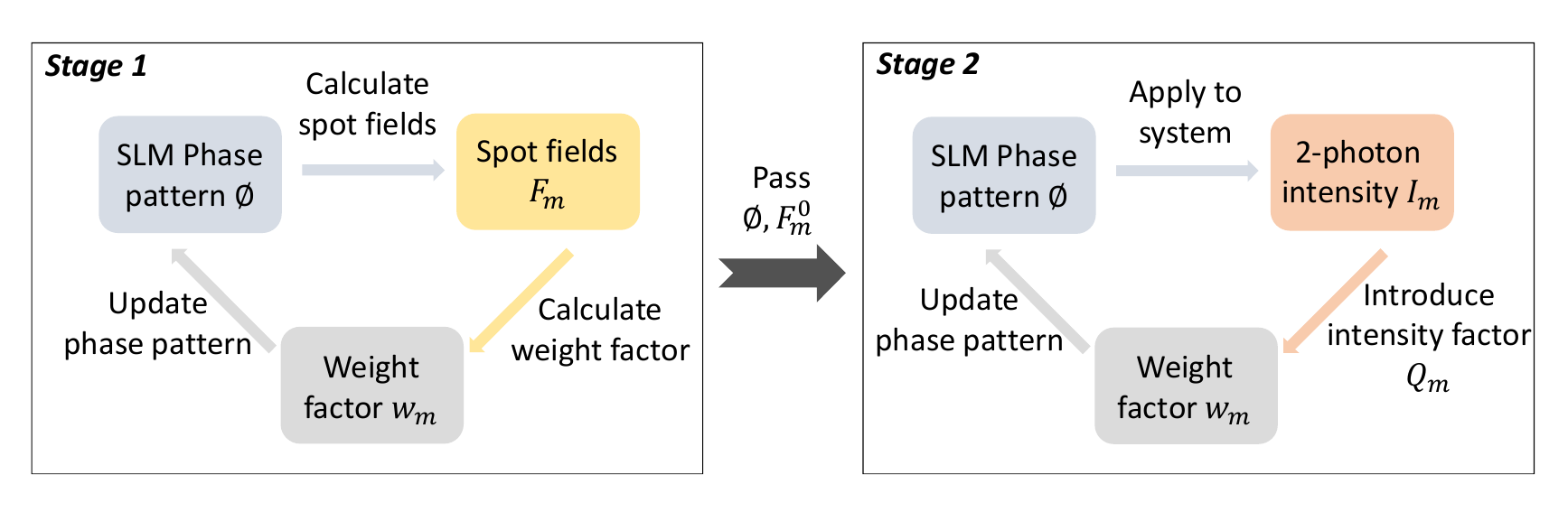}
\caption{Flow chart illustrating the two-stage process for updating the phase pattern in the hologram design.}
\label{chart}
\end{figure}

The phase patterns displayed on the SLM were designed by a Modified Weighted Gerchberg-Saxton method (MWGS)~\cite{DiLeonardo2007, Jesacher2010b}. Two stages of the algorithm were implemented to obtain the final phase pattern, as shown in Figure~\ref{chart}. $Stage~1$ updates the  phase pattern based on theoretical calculation of the intensity in the focal field. The phase pattern in the iteration $p+1$ is updated by,
\begin{equation}\label{phase1}
{\Phi ^{p + 1}} = \mathrm{arg}\left[\sum\limits_{m=1}^n {\left\{ {\exp \left[ {i{\Delta _m}} \right] \times \frac{{F_m^p}}{{\left| {F_m^p} \right|}} \times w_m^p} \right\}}\right]
\end{equation} where weight factor $w_m$ is defined as,
\begin{equation}\label{weight1}
w_m^p = w_m^{p - 1} \times \frac{{\left( {\sum\limits_1^{\rm{n}} {\left\{ {\left| {F_m^p} \right|} \right\}} } \right)/n}}{{\left| {F_m^p} \right|}}
\end{equation} In the above equations, ${\Delta _m}\left( {x,y} \right)$ is the propagator function relating the spatial positions of each spot $m$ in the focal field to pixels of the hologram in  the back focal plane of the lens, compensating for any depth dependent spherical aberration~\cite{Jesacher2010b}. $F_m^p$ is the calculated amplitude of the field for spot $m$ based on the phase in current iteration $p$. $n$ is the total number of spatiotemporal focal spots.

$Stage~1$ designs the initial hologram through pure theoretical simulation. In practice, imperfections of the optical system make it difficult to obtain high quality SSTF as predicted by theory. We therefore introduced a further stage to optimize the hologram. $Stage~2$ of the algorithm updates the phase pattern based on the experimental measurement of two-photon fluorescence from each focus~\cite{Hasagawa2011, SalterSPIE2013}. We also introduced an intensity factor, $Q_m$, which allows a different target intensity for each spot. Such an intensity factor could equally be included in $Stage~1$ of the calculation~\cite{Kaakkunen2013}, but through inclusion in $Stage~2$ it can remove any relation to the non-linearity of the light-matter interaction. The measured two-photon fluorescence intensity of each spatiotemporal focus, in iteration $q$, is represented as $I_m^q$ for spot $m$. The phase pattern in iteration $q+1$ is then updated by,
\begin{equation}\label{phase2}
{\Phi ^{q + 1}} = \mathrm{arg}\left[\sum\limits_1^n {\left\{ {\exp \left[ {i{\Delta _m}} \right] \times \frac{{F_m^0}}{{\left| {F_m^0} \right|}} \times w_m^q} \right\}}\right]
\end{equation} where $F_m^0$ is the last spot field calculated in $Stage~1$. The weight factor $w_m^q$ is defined as,
\begin{equation}\label{weight2}
w_m^q = w_m^{q - 1} \times \frac{{\left( {\sum\limits_1^n {\left\{ {I_m^q/{Q_m}} \right\}} } \right)/n}}{{I_m^q/{Q_m}}}
\end{equation}
As seen in the above equations, in contrast to $Stage~1$, the phase of each spot (which is determined by ${\exp \left[ {i{\Delta _m}} \right]\times\frac{{F_m^0}}{{\left| {F_m^0} \right|}}}$) is fixed in each iteration of $Stage~2$. Only the weight factor $w_m^q$ is updated in $Stage~2$ until all the spots reach the target values for the two-photon fluorescence intensity. In our experiments, 1 to 5 iterations of $Stage~2$ were usually required to obtain a good uniformity, the number of iterations being dependent on the system configuration.

In $Stage~1$ the focal amplitude calculations were based on Fourier optics theory using practical parameters from the experimental system. The simulations of multiple widefield SSTF spots considered both phase modulation and spatial spectral chirp to increase the accuracy of the hologram calculation~\cite{Sun2014}.

\subsection*{Measurement of focal intensity distribution}

Two methods were employed for measuring the focal intensity distribution of the generated spots. In the first, a mirror was aligned at the focal plane of the objective lens and images were taken using the integrated widefield optical microscope. These images were linear in the focal intensity and hence did not reveal the temporal focusing aspect of the arrangement. Therefore, we additionally probed the non-linear response to the SSTF foci by monitoring the fluorescence emission due to two-photon excitation of a uniform fluorescent volume. The sample consisted of an aqueous solution of propidium iodide (PI, Sigma Aldrich P4170) with concentration of 10~mg/mL, sealed between a microscope slide and a cover slip. The emitted fluorescence was captured on the CCD, with an additional short-pass filter inserted to block any back scattered light from the laser. The two-photon nature of the excitation was confirmed by the quadratic dependence of the fluorescence intensity on the illumination power. The axial extent of the foci were estimated by measurement of the edge response obtained by scanning the sample along the optic axis, so that the foci traversed the interface between the cover slip and the PI solution (Supplementary information Figure~S2).

For all of the following results, the system was configured such that each of the SSTF foci had a lateral full width half maximum (FWHM) diameter of 2.5~$\mu$m, which was measured consistently across all experiments. From the two-photon edge response, the axial FWHM in water was estimated to be approximately 7.5~$\mu$m.  These values can be compared with calculations for an objective lens with a NA of 0.2 using conventional focussing: the lateral FWHM would again be 2.5~$\mu$m, whereas the axial FWHM would be around eight times larger at 60~$\mu$m.  The confinement of the non-linear effect along the axial direction clearly shows the non-conventional, spatiotemporal nature of the focus.

\subsection*{Direct laser writing}

A further application that highlights the temporal compression of the pulse in SSTF is direct laser writing inside transparent materials, which relies on non-linear absorption at the focus~\cite{Joglekar2004c}.  Fabrication was conducted either on the surface or inside  standard microscope glass slides (CarlRoth 0656.1). A scan speed of 20~$\mu$m/s was used to draw continuous lines within the glass with a laser repetition rate of 50~kHz. The pulse energy was adjusted as necessary for each experiment.

\section*{RESULTS AND DISCUSSION}

\subsection*{Holographically generated SSTF focal arrays}

\begin{figure}[!h]
\centering\includegraphics[width=17cm]{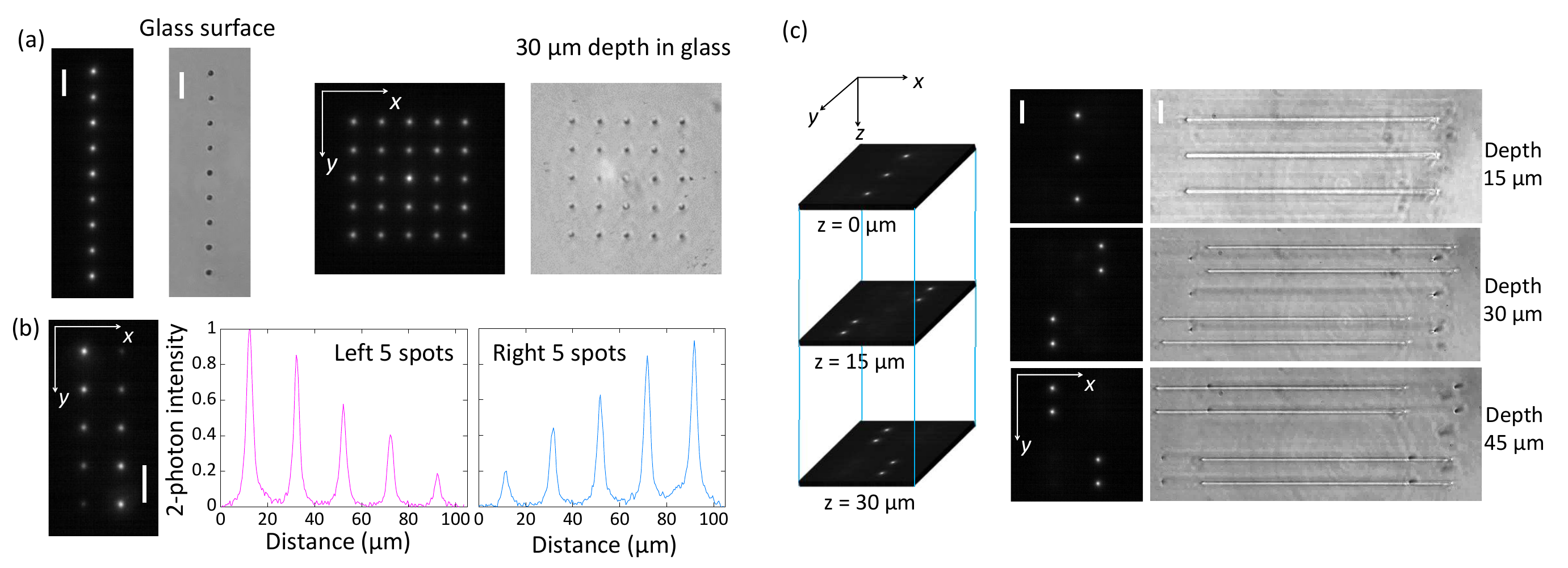}
\caption{(a) The images of two-photon emission and fabrication in glass for a 2D 1$\times$9 array and a 2D 5$\times$5 SSTF multiple foci array. The images from left to right: two-photon image of 1$\times$9 array, single exposure fabrication on the glass surface of 1$\times$9 array, two-photon image of 5$\times$5 array, single exposure fabrication in glass with 30~$\mu$m depth of 5$\times$5 array. Scale bar is 15~$\mu$m. (b) Left: Image of two-photon fluorescence arising from a SSTF array with designed intensity variation between each spot. Scale bar is 20~$\mu$m. Right: The line plots show an intensity profile along the central line of the left and right hand sets of 5 foci. The constant background offset was subtracted in the plots. (c) 3D multiple foci array of 11 spatiotemporal spots which were distributed at three different axial planes. The images of two-photon emission and fabrication in glass are presented. Scale bar is 15~$\mu$m. The SLM phases used to create these temporal arrays are presented in Supplementary information Figure~S3.}
\label{2D3D}
\end{figure}

 Figure~\ref{2D3D} shows various demonstrations of holographically multiplexed spatiotemporal foci in two and three dimensions. Figure~\ref{2D3D}(a) shows images of the two-photon fluorescence and direct laser writing in glass with 1$\times$9 and 5$\times$5 arrays of SSTF foci. The 1$\times$9 SSTF array was designed for fabrication on the glass surface, without the correction of refractive index mismatch, while the 5$\times$5 temporal foci array was designed for fabrication at a 30~$\mu$m depth inside the glass with incorporated  correction of index mismatch aberration. The intensity factor $Q_m$ (Equation~\ref{weight2}) was set to be one to achieve high uniformity for all the SSTF foci. As a result, in both arrays, all the SSTF spots exhibited similar two-photon intensity and similar properties in fabrication. It is notable that the zero order, comprising light that is not modulated by the SLM phase pattern, becomes stronger when the focal array stretches along both the $x$ and $y$ directions. To suppress the intensity of the zero order,  the hologram may be designed to generate an additional  spot which overlaps the zero order and leads to destructive interference~\cite{Jesacher2010b}. It is thus possible to reduce the zero order in the 5$\times$5 array such that the intensity of the central spot is not too significantly different to the others (Supplementary information Figure~S4).

We demonstrate that the intensity of each SSTF spot can also be individually controlled. This relies on the introduction of the intensity factor in $Stage~2$ of the phase pattern design. We designed a 2$\times$5 temporal focal array to demonstrate this ability, as shown in Figure~\ref{2D3D}(b). The intensity factor $Q_m$ (Equation~\ref{weight2}) was set to be [5,~4,~3,~2,~1] and [1,~2,~3,~4,~5] for the spot vectors on the left side and right side, respectively. After a few iterations in $Stage~2$, the two-photon intensity of each spot in the 2$\times$5 array presented the desired relative values. We note that the spot intensities do not necessarily have to be in a linear relationship. Any combination of relative intensities can be achieved by adjusting the intensity factors $Q_m$.

A three dimensional spatiotemporal foci array was designed and presented in Figure~\ref{2D3D}(c). Similarly, the intensity factors $Q_m$ (Equation~\ref{weight2}) were set to one to achieve high uniformity in two-photon intensity. The 11 temporal foci were designed to distribute in three different lateral planes spaced axially by a distance of 15~$\mu$m. In the nominal focal plane, three temporal spots are located along the $y$ direction. In the planes which were 15~$\mu$m and 30~$\mu$m below nominal focal plane, four SSTF spots were distributed in both $x$ and $y$ as shown in Figure~\ref{2D3D}(c). The two-photon fluorescence images in the $z$~=~15~$\mu$m and $z$~=~30~$\mu$m  planes were captured by remotely refocusing the integrated microscope. The refractive index mismatch aberrations were all corrected in the hologram design for 11 SSTF spots, and the array was designed for fabrication in glass at a mean depth of 30~$\mu$m. As shown in the right images of Figure~\ref{2D3D}(c), with one single pass of the glass sample, 11 lines with different starting points were fabricated at three different depths at the same time.

In the design of a 3D SSTF focal array, it is notable that if several spots have the same $x$ and $y$ positions, the spots had to have considerable separation along the $z$ direction (at least 5 - 10 times of the FWHM axial resolution) in order to achieve distinct temporal foci with high  uniformity of intensity. The required separation is related to the monochromatic beam size in the objective pupil (i.e. the NA for each spectral component) and hence the transverse size of the foci. We note that this distance is greater than that needed in holographic arrays using conventional focusing, due to the reduced effective NA of each spectral component in SSTF.

\subsection*{Interleaving foci in space and time}

We have shown that it is possible to generate multiple SSTF spots in 3D, using a hologram in the spatially chirped beam. We now explore how additional consideration of the time dimension leads to extra flexibility for implementation of parallelized SSTF.

\begin{figure}[htb]
\centering\includegraphics[width=8cm]{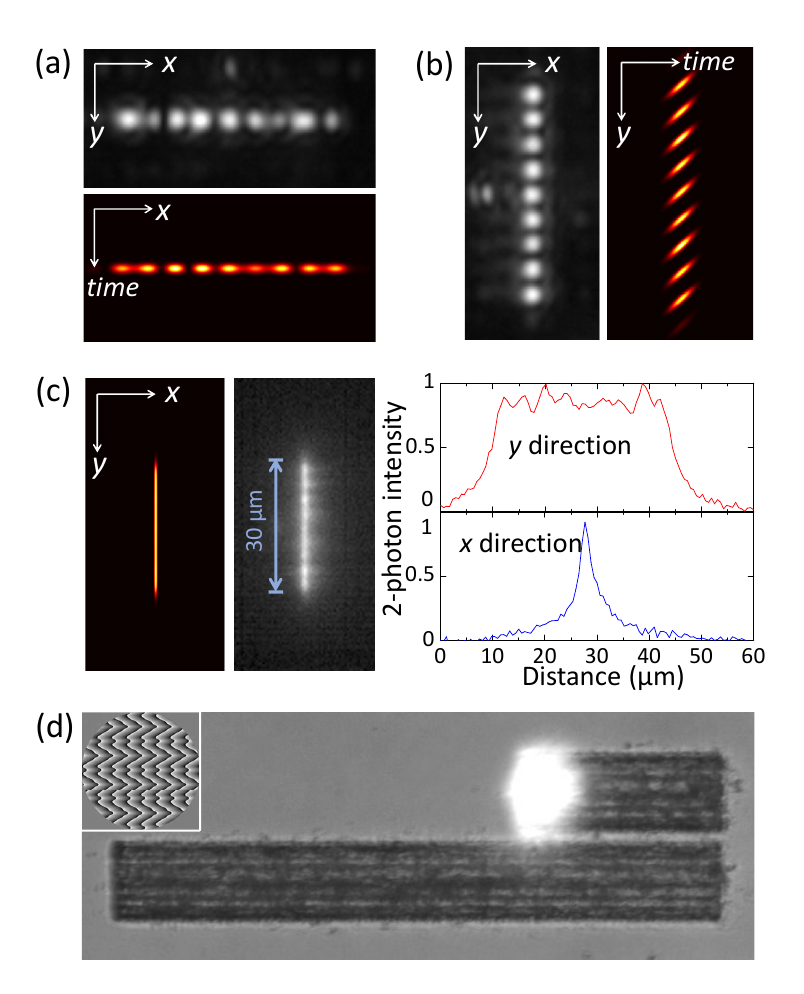}
\caption{(a) Top: Image of nine spatiotemporal spots in an array orthogonal to the direction of spatial-chirp ($x$ axis). Bottom: simulation of intensity profile in the time domain ($x-time$ profile). The distance between each spot is 4~$\mu$m. (b) Left: Image of nine spatiotemporal spots array along the direction of spatial-chirp $y$ axis). Right: simulation of intensity profile in time domain ($y-time$ profile). The distance between each spot is 4~$\mu$m. (c) Left: simulation of time-averaged two-photon intensity for line temporal focusing with nine spots, assuming the spots are separated by a distance of 3~$\mu$m. Middle: experimental two-photon image taken in the fluorescence dye sample. Right: plots of the two-photon intensities along the $x$ and $y$ axis. (d) The application of the uniform intensity temporal line to direct laser writing on the surface of glass. The focal plasma can be seen as the white region. Inset: phase pattern used to create this uniform intensity temporal line.}
\label{inteference}
\end{figure}

We first consider a linear array of nine SSTF spots arranged along the direction perpendicular to the spatial chirp (i.e. parallel to the $x$ axis) with a separation of 4~$\mu$m. The  experimentally measured spatial intensity distributions are shown in Figure~\ref{inteference}(a) along with the theoretically calculated space time distribution. As expected, there is mutual interference between the spots, and the definition of the spot array is lost, as the light arrives simultaneously at each point across the array. However, if we create another array, same in spacing but oriented along the spatial chirp direction ($y$ axis) there is minimal interference and the array of nine spots is still clearly defined (Figure~\ref{inteference}(b)). The reason for this lack of interference can be explained by the strong pulse front tilt (PFT) present in SSTF foci~\cite{Vitek2010,Block2014,Zhang2014}. The space-time plot shows clearly that due to the PFT as there is no overlap between the foci when considered in 4D.  Instead the foci are interleaved in space-time and there is no interference between the spots.

This space-time interleaving means that it is possible to generate an extended line of approximately uniform time-averaged intensity using multiple SSTF spots positioned along the direction of the spatial chirp. This can be realized either by reducing the distance between each spot, or increasing the focal spot size through reducing the input monochromatic beam size. In this way, adjacent spots can be spatially overlapped with minimal interference between them, as demonstrated by the simulation results in Figure~\ref{inteference}(c) (more simulations in Supplementary information Figure~S6). In practice, the spot spacing was reduced to 3~$\mu$m, and the focal spots were made broader by the inclusion of an adjustable iris before the gratings.  Figure~\ref{inteference}(c) presents the two-photon fluorescence generated using the focal strip, which extends along the spatial-chirp direction ($y$ axis) for more than 30~$\mu$m.  This remarkable result, where it is possible to spatially overlap several foci from the same pulse is only possible due to spatiotemporal interleaving at the focus. It is not possible to generate such a uniform line focus in the orthogonal direction, in which there is no PFT.

One application of this uniform intensity line focus is in wide area laser machining, where ``flat top" beams are commonly sought. As shown in Figure~\ref{inteference}~(d), a region with size 30$\times$200~$\mu$m was fabricated with a single pass of the sample. We note that laser processing  with a conventional line focus generated by, for example, a cylindrical lens, tends to produce stronger material modification in the central area, where the laser intensity is higher. In contrast, the spatiotemporal interleaving method generates an intensity distribution with a flatter top leading to a fabricated region which is uniform across the whole area, maximizing processing efficiency.  Further applications of such uniform line foci could be found in parallel multiphoton microscopy.

\section*{CONCLUSION}

We have demonstrated that multiple SSTF foci may be generated in 2D and 3D through an appropriately designed diffractive holographic element illuminated by the spatially chirped beam. The spatiotemporal nature of these foci has been confirmed through two-photon imaging and direct laser writing. Thus, it is possible to combine the advantageous aspects of SSTF with the ability to  increase process speed through parallelization. In addition, we discovered an unique characteristic of multiplexed SSTF foci that the spatio-temporal coupling may be exploited to interleave foci in space-time.  This permits spatial overlap of adjacent foci with minimal interference.  It is thus possible to generate a uniform ``flat top" intensity extended line focus based upon the inherent pulse front tilt present in SSTF. This ability offers a further degree of freedom over conventional holographic spot arrays with conventional focusing, where there is a limitation on how closely spaced the spots may be~\cite{Hasagawa2012}. If they are positioned closer than around 3 Airy units of each other, mutual interference causes significant distortion of the desired intensity profile, in effect resulting in non-uniformity in the illuminated area.  Appropriate placement of the foci in 4D allows us to avoid this problem, thus opening up applications in non-linear imaging or laser processing.

\section*{ACKNOWLEDGMENTS}

The authors gratefully acknowledge experimental guidance and useful discussion with Marcus Baum and Tom Hafner. The research was supported through funding from the Erlangen Graduate School in Advanced Optical Technologies (SAOT) by the German Research Foundation (DFG) in the framework of the German excellence initiative, the Leverhulme Trust (RPG-2013-044) and by the European Research Council (ERC) under the Horizon 2020 research and innovation programme (grant agreement no. 695140).


\begin{thebibliography}{10}
\newcommand{\enquote}[1]{``#1''}

\bibitem{ritsch2011}
C.~Maurer, S.~Bernet, A.~Jesacher, and M.~Ritsch-Marte, \enquote{{What spatial
  light modulators can do for optical microscopy},} Laser Photonics Reviews
  \textbf{5}, 81--101 (2011).

\bibitem{grier2002}
J.~E. Curtis, B.~A. Koss, and D.~G. Grier, \enquote{{Dynamic holographic
  optical tweezers},} Opt. Comms \textbf{207}, 169--175 (2002).

\bibitem{Sugioka2014}
K.~Sugioka and Y.~Cheng, \enquote{{Femtosecond laser three-dimensional micro-
  and nanofabrication},} Appl. Phys. Rev \textbf{1}, 041303 (2014).

\bibitem{preece2009}
D.~Preece, R.~Bowman, A.~Linnenberger, G.~Gibson, D.~Serati, and M.~Padgett,
  \enquote{{Increasing trap stiffness with position clamping in holographic
  optical tweezers},} Opt. Express \textbf{17}, 22718--22725 (2009).

\bibitem{xu05}
G.~Zhu, J.~van Howe, M.~Durst, W.~Zipfel, and C.~Xu, \enquote{{Simultaneous
  spatial and temporal focusing of femtosecond pulses},} Opt. Express
  \textbf{13}, 2153--2159 (2005).

\bibitem{Oron2005}
D.~Oron and Y.~Silberberg, \enquote{{Spatiotemporal coherent control using
  shaped, temporally focused pulses},} Optics Express \textbf{13}, 9903--9908
  (2005).

\bibitem{Durfee2014}
C.~G. Durfee and J.~A. Squier, \enquote{{Breakthroughs in Photonics 2014:
  Spatiotemporal Focusing: Advances and Applications},} IEEE Photonics Journal
  \textbf{7}, 0700806 (2014).

\bibitem{Choi2013}
H.~Choi, E.~Y.~S. Yew, B.~Hallacoglu, S.~Fantini, C.~J.~R. Sheppard, and
  P.~T.~C. So, \enquote{{Improvement of axial resolution and contrast in
  temporally focused widefield two-photon microscopy with structured light
  illumination},} Biomed. Opt. Express \textbf{4}, 995--1005 (2013).

\bibitem{Lien2014}
C.-H. Lien, C.-Y. Lin, S.-J. Chen, and F.-C. Chien, \enquote{{Dynamic particle
  tracking via temporal focusing multiphoton microscopy with astigmatism
  imaging},} Opt. Express \textbf{22}, 27290--27299 (2014).

\bibitem{Dana2014}
H.~Dana, A.~Marom, S.~Paluch, R.~Dvorkin, I.~Brosh, and S.~Shoham,
  \enquote{{Hybrid multiphoton volumetric functional imaging of large-scale
  bioengineered neuronal networks},} Nat. Commun. \textbf{5}, 3997 (2014).

\bibitem{Song2015}
Q.~Song, A.~Nakamura, K.~Hirosawa, K.~Isobe, K.~Midorikawa, and F.~Kannari,
  \enquote{{Two-dimensional spatiotemporal focusing of femtosecond pulses and
  its applications in microscopy},} Rev. Sci. Inst. \textbf{86}, 083701 (2015).

\bibitem{Shoham2013}
H.~Dana, N.~Kruger, A.~Ellman, and S.~Shoham, \enquote{{Line temporal focusing
  characteristics in transparent and scattering media},} Opt. Express
  \textbf{21}, 5677--5687 (2013).

\bibitem{Vitek2010a}
D.~N. Vitek, D.~E. Adams, A.~Johnson, P.~S. Tsai, S.~Backus, C.~G. Durfee,
  D.~Kleinfeld, and J.~A. Squier, \enquote{{Temporally focused femtosecond
  laser pulses for low numerical aperture micromachining through optically
  transparent materials},} Optics express \textbf{18}, 18086--94 (2010).

\bibitem{Kammel2014}
R.~Kammel, R.~Ackermann, J.~Thomas, J.~G{\"{o}}tte, S.~Skupin,
  A.~T{\"{u}}nnermann, and S.~Nolte, \enquote{{Enhancing precision in fs-laser
  material processing by simultaneous spatial and temporal focusing},} Light:
  Science {\&} Applications \textbf{3}, e169 1--8 (2014).

\bibitem{He2010}
F.~He, H.~Xu, Y.~Cheng, J.~Ni, H.~Xiong, Z.~Xu, K.~Sugioka, and K.~Midorikawa,
  \enquote{{Fabrication of microfluidic channels with a circular cross section
  using spatiotemporally focused femtosecond laser pulses},} Opt. Lett.
  \textbf{35}, 1106--1108 (2010).

\bibitem{He2011}
F.~He, Y.~Cheng, J.~Lin, J.~Ni, Z.~Xu, K.~Sugioka, and K.~Midorikawa,
  \enquote{{Independent control of aspect ratios in the axial and lateral cross
  sections of a focal spot for three-dimensional femtosecond laser
  micromachining},} New Journal of Physics \textbf{13} (2011).

\bibitem{Vitek2010}
D.~N. Vitek, E.~Block, Y.~Bellouard, D.~E. Adams, S.~Backus, D.~Kleinfeld,
  C.~G. Durfee, J.~A. Squier, A.~Johnson, and P.~S. Tsai,
  \enquote{{Spatio-temporally focused femtosecond laser pulses for
  nonreciprocal writing in optically transparent materials},} Opt. Express
  \textbf{18}, 24673--24678 (2010).

\bibitem{Block2014}
E.~Block, J.~Thomas, C.~Durfee, and J.~Squier, \enquote{{Integrated single
  grating compressor for variable pulse front tilt in simultaneously spatially
  and temporally focused systems.}} Optics letters \textbf{39}, 6915--6918
  (2014).

\bibitem{Zhang2014}
S.~Zhang, F.~Wyrowski, R.~Kammel, and S.~Nolte, \enquote{{A brief analysis on
  pulse front tilt in simultaneous spatial and temporal focusing},} J. Opt.
  Soc. Am. A \textbf{31}, 2437--2446 (2014).

\bibitem{Stoian2013}
R.~Stoian, J.~P. Colombier, C.~Mauclair, G.~Cheng, M.~K. Bhuyan, P.~K. Velpula,
  and P.~Srisungsitthisunti, \enquote{{Spatial and temporal laser pulse design
  for material processing on ultrafast scales},} Applied Physics A
  \textbf{114}, 119--127 (2013).

\bibitem{Poland2014}
S.~P. Poland, N.~Krstajic, S.~Coelho, D.~Tyndall, R.~J. Walker, V.~Devauges,
  P.~E. Morton, N.~S. Nicholas, J.~Richardson, D.~D.-U. Li, K.~Suhling, C.~M.
  Wells, M.~Parsons, R.~K. Henderson, and S.~M. Ameer-Beg,
  \enquote{{Time-resolved multifocal multiphoton microscope for high speed FRET
  imaging in vivo},} Opt. Lett. \textbf{39}, 6013--6016 (2014).

\bibitem{Waller2013}
E.~H. Waller and G.~von Freymann, \enquote{{Multi foci with diffraction limited
  resolution},} Opt. Express \textbf{21}, 21709 (2013).

\bibitem{Lin2013}
H.~Lin and M.~Gu, \enquote{{Creation of diffraction-limited non-Airy multifocal
  arrays using a spatially shifted vortex beam},} Appl. Phys. Lett.
  \textbf{102}, 084103 (2013).

\bibitem{Hasagawa2016}
S.~Hasegawa, H.~Ito, H.~Toyoda, and Y.~Hayasaki, \enquote{{Massively parallel
  femtosecond laser processing},} Opt. Express \textbf{24}, 18513 (2016).

\bibitem{Papagiakoumou2013}
E.~Papagiakoumou, A.~B{\`{e}}gue, B.~Leshem, O.~Schwartz, B.~M. Stell,
  J.~Bradley, D.~Oron, and V.~Emiliani, \enquote{{Functional patterned
  multiphoton excitation deep inside scattering tissue},} Nature Photonics
  \textbf{7}, 274--278 (2013).

\bibitem{Hernandez2016}
O.~Hernandez, E.~Papagiakoumou, D.~Tanesee, K.~Felin, C.~Wyart, and
  V.~Emiliani, \enquote{{Three-dimensional spatiotemporal focusing of
  holographic patterns},} Nature Communications \textbf{7}, 11928 (2016).

\bibitem{Papagiakoumou2008}
E.~Papagiakoumou, V.~de~Sars, D.~Oron, and V.~Emiliani, \enquote{{Patterned
  two-photon illumination by spatiotemporal shaping of ultrashort pulses},}
  Optics Express \textbf{16}, 22039--22047 (2008).

\bibitem{Kim2010}
D.~Kim and P.~T.~C. So, \enquote{{High-throughput three-dimensional
  lithographic microfabrication},} Optics Letters \textbf{35}, 1602--1604
  (2010).

\bibitem{Li2012}
Y.-C. Li, L.-C. Cheng, C.-Y. Chang, C.-H. Lien, P.~J. Campagnola, and S.-J.
  Chen, \enquote{{Fast multiphoton microfabrication of freeform polymer
  microstructures by spatiotemporal focusing and patterned excitation},} Optics
  express \textbf{20}, 19030--19038 (2012).

\bibitem{Papagiakoumou2010}
E.~Papagiakoumou, F.~Anselmi, A.~Begue, V.~De~Sars, J.~Gluckstad, E.~Y.
  Isacoff, and V.~Emiliani, \enquote{{Scanless two-photon excitation of
  channelrhodopsin-2},} Nat. Methods \textbf{7}, 848--854 (2010).

\bibitem{Durst2006}
M.~E. Durst, G.~Zhu, and C.~Xu, \enquote{{Simultaneous spatial and temporal
  focusing for axial scanning},} Optics express \textbf{14}, 12243--54 (2006).

\bibitem{Lesham2014}
B.~Leshem, O.~Hernandez, E.~Papagiakoumou, V.~Emiliani, and D.~Oron,
  \enquote{{When can temporally focused excitation be axially shifted by
  dispersion?}} Opt. Express \textbf{22}, 7087 (2014).

\bibitem{Zurauskas2017}
M.~Zurauskas, O.~Barnstedt, M.~Frade-Rodriguez, S.~Waddell, and M.~J. Booth,
  \enquote{{Rapid Sensing Of Volumetric Neural Activity Through Adaptive Remote
  Focusing},} bioRxiv p. 125070 (2017).

\bibitem{Sun2014}
B.~Sun, P.~S. Salter, and M.~J. Booth, \enquote{{Effects of aberrations in
  spatiotemporal focusing of ultrashort laser pulses},} J. Opt. Soc. Am. A
  \textbf{31}, 765 (2014).

\bibitem{Chang2014}
C.-Y. Chang, L.-C. Cheng, H.-W. Su, Y.~Y. Hu, K.-C. Cho, W.-C. Yen, C.~Xu,
  C.~Y. Dong, and S.-J. Chen, \enquote{{Wavefront sensorless adaptive optics
  temporal focusing-based multiphoton microscopy.}} Biomedical Optics Express
  \textbf{5}, 1768--1777 (2014).

\bibitem{Greco2016}
M.~J. Greco, E.~Block, A.~K. Meier, A.~Beaman, S.~Cooper, M.~Iliev, J.~A.
  Squier, and C.~G. Durfee, \enquote{{Spatial–spectral characterization of
  focused spatially chirped broadband laser beams},} Appl. Opt. \textbf{54},
  9818--9822 (2015).

\bibitem{Durfee2012}
C.~G. Durfee, M.~J. Greco, E.~Block, D.~Vitek, and J.~A. Squier,
  \enquote{{Intuitive analysis of space-time focusing with double-ABCD
  calculation},} Opt. Express \textbf{20}, 14244 (2012).

\bibitem{DiLeonardo2007}
R.~{Di Leonardo}, F.~Ianni, and G.~Ruocco, \enquote{{Computer generation of
  optimal holograms for optical trap arrays},} Optics express \textbf{15},
  1913--1922 (2007).

\bibitem{Jesacher2010b}
A.~Jesacher and M.~J. Booth, \enquote{{Parallel direct laser writing in three
  dimensions with spatially dependent aberration correction},} Opt. Express
  \textbf{18}, 21090--21099 (2010).

\bibitem{Hasagawa2011}
S.~Hasegawa and Y.~Hayasaki, \enquote{{Second-harmonic optimization of
  computer-generated hologram},} Opt. Lett. \textbf{36}, 2943 (2011).

\bibitem{SalterSPIE2013}
P.~S. Salter and M.~J. Booth, \enquote{{Dynamic optical methods for direct
  laser written waveguides},} SPIE MOEMS-MEMS p. 86130A (2013).

\bibitem{Kaakkunen2013}
M.~Silvennoinen, J.~Kaakkunen, K.~Paivasaari, and P.~Vahimaa,
  \enquote{{Parallel femtosecond laser ablation with individually controlled
  intensity},} Opt. Express \textbf{22}, 2603 (2014).

\bibitem{Joglekar2004c}
A.~P. Joglekar, H.-H. Liu, E.~Meyh{\"{o}}fer, G.~Mourou, and A.~J. Hunt,
  \enquote{{Optics at critical intensity: applications to nanomorphing},} P.
  Natl. Acad. Sci. USA \textbf{101}, 5856--61 (2004).

\bibitem{Hasagawa2012}
Y.~Hayasaki, M.~Nishitani, H.~Takahashi, A.~Yamamoto, H~Takita, D.~Suzuki, and
  S.~Hasegawa, \enquote{{Experimental investigation of the closest parallel
  pulses in holographic femtosecond laser processing},} Appl. Phys. A
  \textbf{107}, 357--362 (2012).

\end{thebibliography}
\bibliographystyle{osajnl}

\end{document}